\documentclass{appolb}
\usepackage{epsfig}
\usepackage{subfigure}
\newcommand{\dd}{\mbox{\rm d}}

\begin{document}
\date{\today}
\pagestyle{plain}
\newcount\eLiNe\eLiNe=\inputlineno\advance\eLiNe by -1
\title{Coupled-channel effects in kaon pair production%
\thanks{Presented at the Symposium on Meson Physics, Cracow 2008}%
}
\author{Colin Wilkin
\address{Physics and Astronomy Dept., UCL, Gower Street, London, WC1E
6BT, UK }} \maketitle

\begin{abstract}
Two coupled channel effects connected with kaon pair production in
proton-proton collisions are discussed. (1) Although there is ample
evidence that the antikaon is strongly attracted to the recoil
protons in $pp\to K^+p\,\{K^-p\}$, residual effects of the $K^+K^-$
interaction are seen, including a possible cusp at the $K^0\bar{K}^0$
threshold. This is investigated within a simple $K$-matrix approach.
(2) The production rates and invariant mass distributions for $pp\to
K^+p\,\{K^-p\}$ and $pp\to K^+p\,\{\pi^0\Sigma^0\}$ are related using
a separable potential description of the coupled $K^-p/\pi^0\Sigma^0$
channels. It can be plausibly argued that this pair of reactions is
driven through the production of the $\Lambda(1405)$.
\end{abstract}

\begin{figure}[htb]
\begin{center}
\includegraphics[clip,width=6cm]{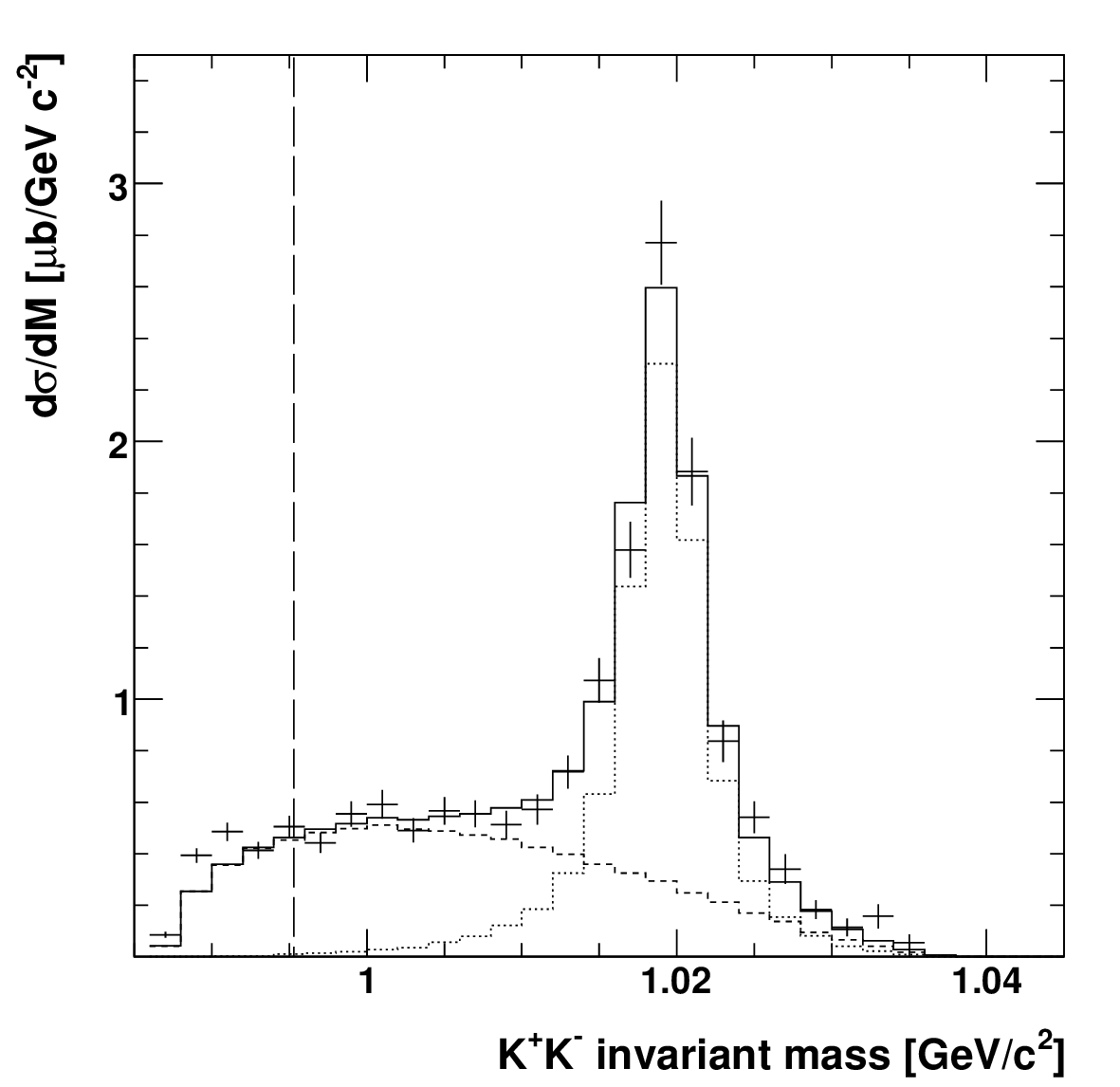}
\includegraphics[clip,width=6.2cm]{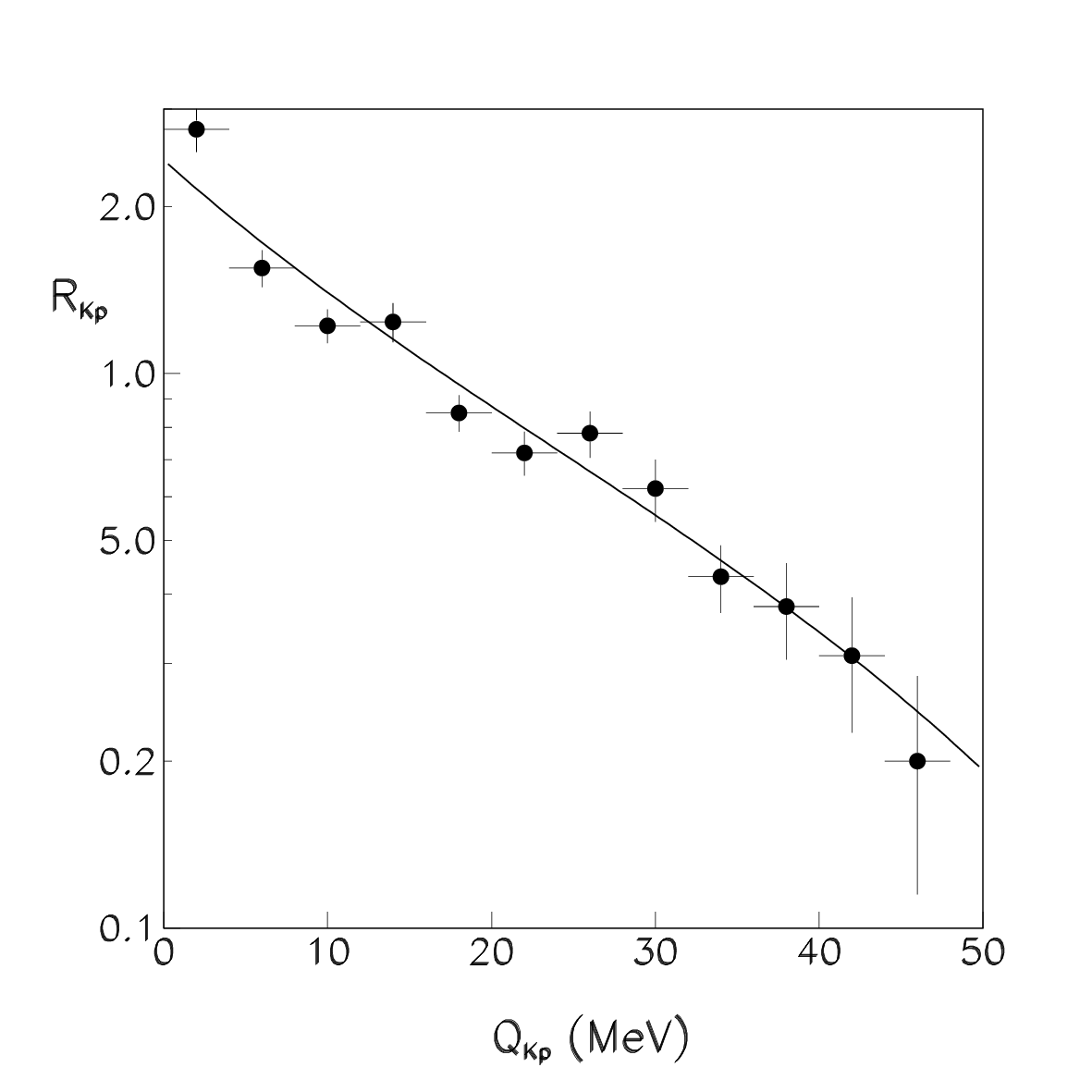}
\caption{Left: Differential cross section for $pp\to ppK^+K^-$ at
2.65~GeV (crosses) as a function of the $K^+K^-$ invariant mass compared to
simulations of the $\phi$ (dotted) and non-$\phi$ (dashed)
contributions and their sum (solid histogram). The $K^0\bar{K}^0$ threshold
is indicated by the dashed vertical line. Right: The ratio of
differential cross sections with respect to $Q_{Kp}=m_{Kp}-m_K-m_p$;
see Eq.~(\ref{ratio_def}). The curve results from the amplitude analysis of
Ref.~\cite{Alexey2}, which includes a $\bar{K}^0d$ \emph{fsi}.
} \label{KKmass}
\end{center}
\end{figure}

The bulk of the observed distributions in $pp\to ppK^+K^-$ above and
below the $\phi$ threshold can be understood in terms of $pp$ and
$K^-p$ final state interactions~\cite{Winter,Maeda08}, as can be seen
from Fig.~\ref{KKmass}. It is shown there that the $K^-p$ \emph{fsi}
distorts particularly the ratio of the differential cross sections
\begin{equation}
\label{ratio_def}%
R_{Kp}=\frac{\dd\sigma/\dd M_{K^-p}}{\dd\sigma/\dd M_{K^+p}}\,, %
\end{equation}
which has a very strong preference for low $Kp$ invariant masses,
$M_{Kp}$.

Since a full treatment of the dynamics of the four-body $ppK^+K^-$
channel is currently impractical, the final state interactions were
introduced in an \emph{ad hoc} way, as the product of the
enhancements in the $pp$ and the two $K^-p$ combinations, all evaluated 
at the appropriate relative momenta $q$~\cite{Maeda08}:
\begin{equation}
\label{assumption} F = F_{pp}(q_{pp})\times F_{Kp}(q_{Kp_1}) \times
F_{Kp}(q_{Kp_2})\,.
\end{equation}
This was used to generate the simulations shown in Fig.~\ref{KKmass}.
The $K^-p$ \emph{fsi} was taken in the scattering length
approximation, $F_{Kp}(q)\approx 1/(1-iqa)$, where $|a|\approx
1.5$~fm. With this value of $a$, the approach reproduces the
$K^-p/K^+p$ ratio also at other energies~\cite{Maeda08}, as well as
the COSY-11 results~\cite{Winter}. Moreover, the simulation suggests
that the $K^-pp$ system should be enhanced at low masses and this
feature is also seen in the ANKE data~\cite{Maeda08}. Further
evidence that the antikaon is attracted to nucleons is to be found in
the $pp\to dK^+\bar{K}^0$ reaction, where low $\bar{K}^0d$ masses are
favoured compared to $K^+d$~\cite{Alexey2}.

\begin{figure}[htb]
\begin{center}
\includegraphics[clip,width=6cm]{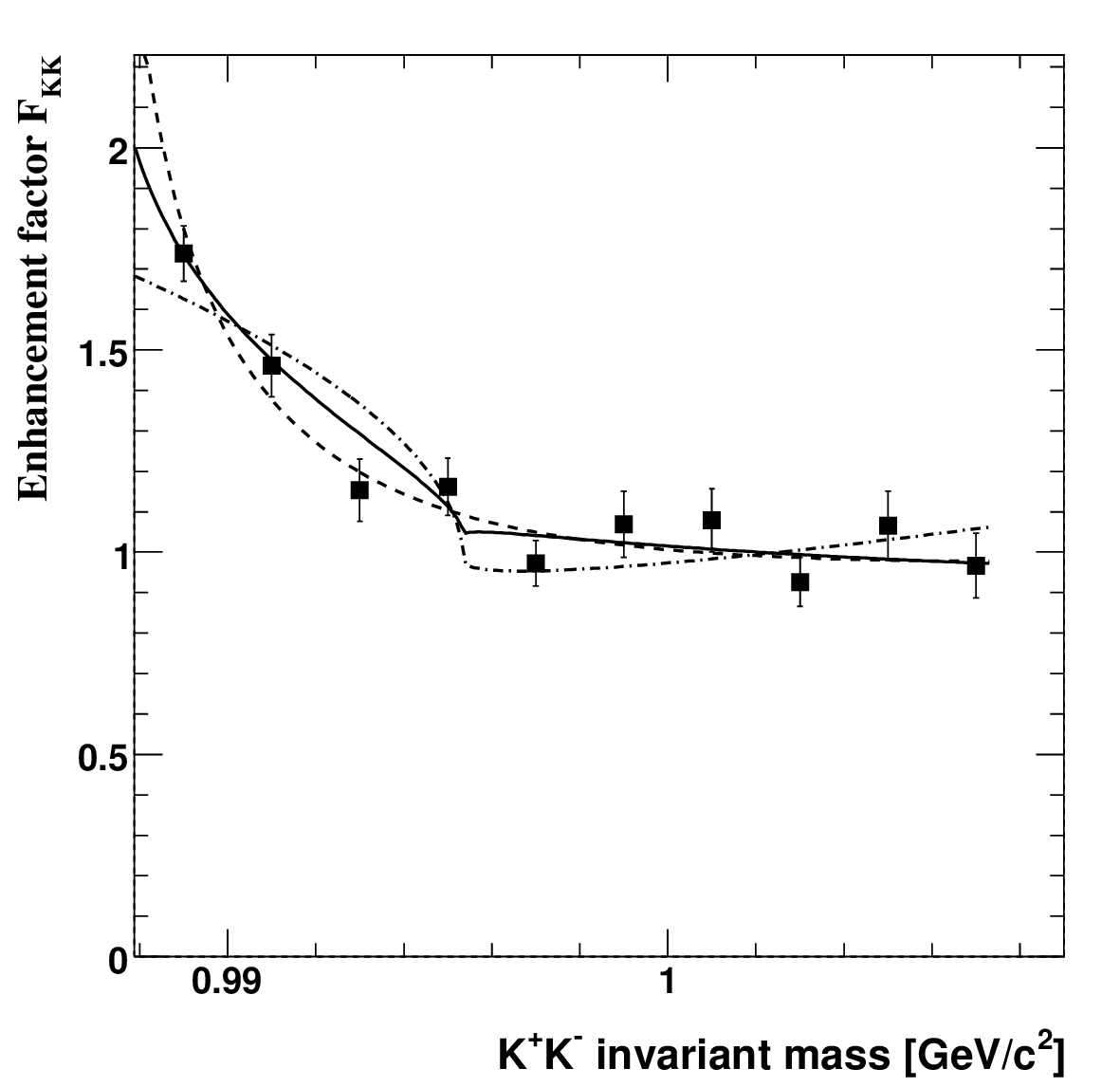}
\caption{Ratio of the $K^+K^-$ invariant mass spectra from the $pp\to
ppK^+K^-$ reaction to the simulation presented in
Ref.~\cite{Maeda08}. The experimental points correspond to the
weighted average of data taken at 2.65, 2.70, and 2.83~GeV. The solid
curve is the result of a best fit of Eq.~(\ref{amps}) to these data.
The dot-dashed curve is the best fit when the elastic rescattering is
\emph{arbitrarily} neglected and the dashed when the charge-exchange
term is omitted. \label{Fit}}
\end{center}
\end{figure}

However, this approach underestimates the data at low $K^+K^-$ masses
in Fig.~\ref{KKmass} and so the \emph{ansatz} of
Eq.~(\ref{assumption}) has to be generalised to include an \emph{fsi}
in this channel. The effects are smaller here and, to illustrate
them, the experimental data at all three ANKE energies have been
divided by the simulations generated by Eq.~(\ref{assumption}), and
their average is plotted in Fig.~\ref{Fit}.

The enhancement seems to be most prominent between the $K^+K^-$ and
$K^0\bar{K}^0$ thresholds at 987.4 and 995.3~MeV/c$^2$, respectively.
It is therefore natural to speculate that it is also influenced by
virtual $K^0\bar{K}^0$ production and its subsequent conversion into
$K^+K^-$ through a charge-exchange \emph{fsi}. If the $s$-wave
$K^+K^-\rightleftharpoons K^0\bar{K}^0$ coupling is strong, this
would generate an observable cusp at the $K^0\bar{K}^0$ threshold.
These possibilities were examined in Ref.~\cite{Alexey3}, where it
was shown that the enhancement factor has a momentum dependence of
the form
\begin{equation}
\mathcal{F}=\left|\frac{B_1/(B_1+B_0)}{\left(1-i\frac{1}{2}q[A_1-A_0]\right)(1-ikA_1)}
+\frac{B_0/(B_1+B_0)}{\left(1-i\frac{1}{2}q[A_0-A_1]\right)(1-ikA_0)}\right|^{\,2}\!.
\label{amps}
\end{equation}

Here $B_0$ and $B_1$ are the \emph{bare} $pp\to ppK\bar{K}$
amplitudes for producing $s$-wave $K\bar{K}$ pairs in isospin-0 and 1
states, respectively. These amplitudes, which already include the
\emph{fsi} in the $K^-p$ and $pp$ channels~\cite{Maeda08}, are then
distorted through a \emph{fsi} corresponding to elastic $K^+K^-$
scattering. This leads to enhancement factors of the form
$1/(1-ikA_I)$, where $k$ is the momentum in the $K^+K^-$ system and
$A_I$ is the $s$-wave scattering length in each of the two isospin
channels. The charge-exchange \emph{fsi} depends upon the
$K^0\bar{K}^0\to K^+K^-$ scattering length, which is proportional to
the difference between $A_0$ and $A_1$, and on the momentum $q$ in
the $K^0\bar{K}^0$ system.

A cusp structure might arise because $q$ changes from being purely
real above the $K^0\bar{K}^0$ threshold to purely imaginary below
this point. The strength of the effect depends upon $A_0-A_1$, but
its shape also depends upon the interference with the direct $K^+K^-$
production amplitude.

There is great uncertainty in the values of the scattering lengths
and the choices made in Ref.~\cite{Alexey3},
$A_1=(0.1\pm0.1)+i(0.7\pm0.1)$~fm and
$A_0=(-0.45\pm0.2)+i(1.63\pm0.2)$~fm, imply a significant
charge-exchange contribution. The subsequent fitting of the data on
the basis of Eq.~(\ref{amps}) is best achieved with $|B_1/B_0|^2 =
0.38_{-0.14}^{+0.24}$, \emph{i.e.}, the kaon pairs are produced
dominantly in the isospin-zero combination.

The resulting fit shown in Fig.~\ref{Fit} manifests a cusp at the
$K^0\bar{K}^0$ threshold, though the data themselves are not
sufficiently precise to see this unambiguously. The other fits shown
there are non-allowed solutions, where one neglects either the
elastic or charge-exchange \emph{fsi}.

\begin{figure}[h]
\begin{center}
\includegraphics[clip,width=6cm, angle=0]{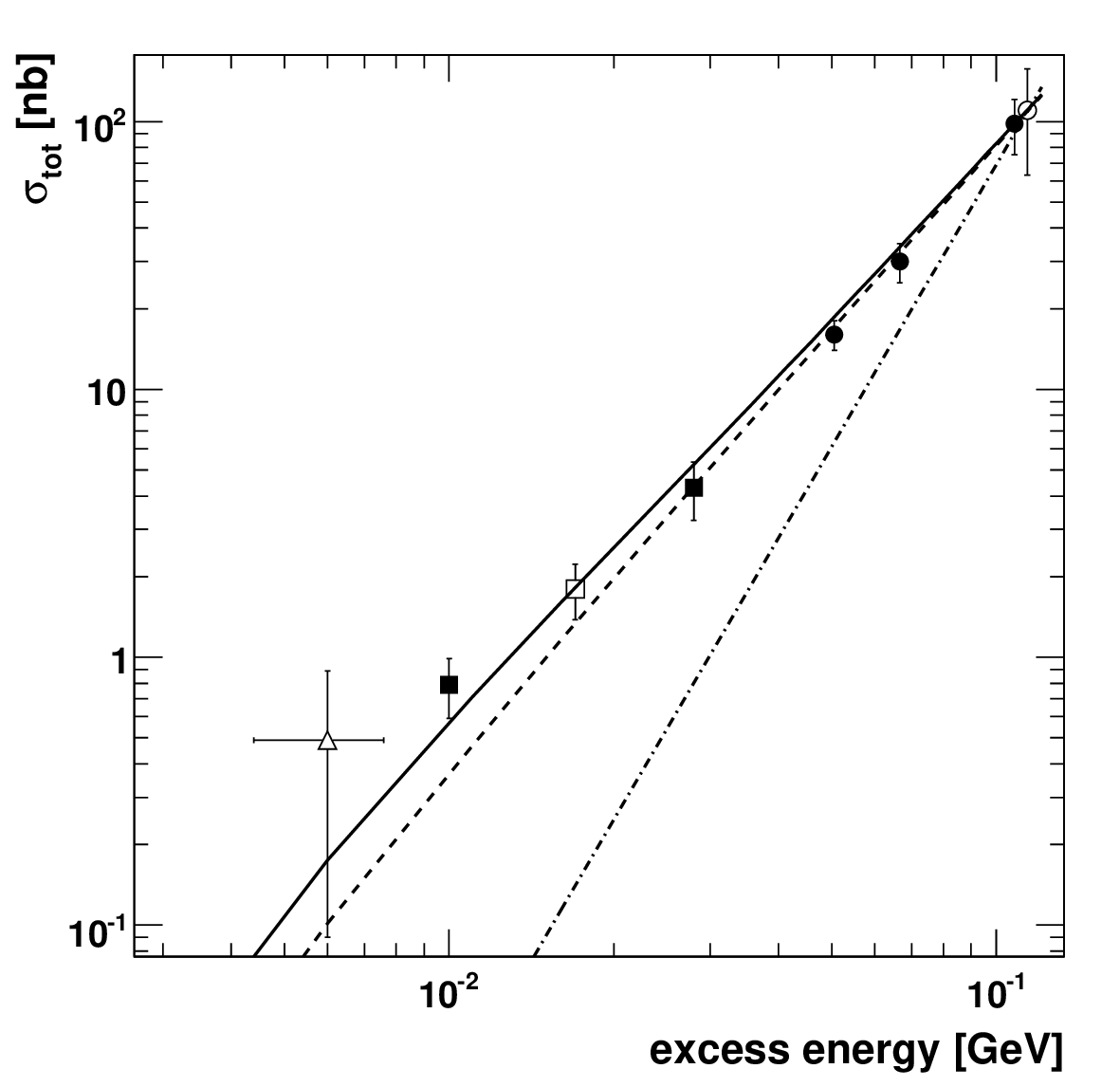}
\caption{Experimental total cross sections for $pp\to ppK^+K^-$ as a
function of the excess energy. The dot-dashed curve is that of
four-body phase space normalised on the 108~MeV point. The dashed
curve includes final state interactions between the $K^-$ and the
protons and between the two protons themselves~\cite{Maeda08}. The
further consideration of the \emph{fsi} between the kaons leads to
the solid curve~\cite{Alexey3}. \label{sigt}}
\end{center}
\end{figure}

The energy dependence of the total cross section shown in
Fig.~\ref{sigt} is definitely improved when the $K\bar{K}$ \emph{fsi}
is included but these, and especially the differential data, have to
be improved in order to be compelling.

In addition to the $pp\to pK^+\{pK^-\}$ measurement, the ANKE
collaboration also extracted data on the $pp\to
pK^+\{\Sigma^0\pi^0\}$ reaction at 3.65~GeV/$c$~\cite{Iza}. As a
second example of a coupled-channel effect, I would like to argue
that both data sets might be understood in terms of the production
and decay of the $\Lambda(1405)$, even though this resonance has a
nominal mass below the $pK^-$ threshold. To investigate this we have
to study the coupled $K^-p \rightleftharpoons\pi^0\Sigma^0$ systems
in some detail. This is easiest to achieve within the realm of a
separable potential description because the resulting equations can
be solved algebraically. Separate and conquer~\cite{Brown}\,!

A separable description of the $I=0$ coupled--channel system has been
given in Ref.~\cite{Gal}. Here the potential is taken in the form
\begin{equation}
\label{Yamaguchi} V_{ij}(p,p') =
(2\pi)^3\,\frac{A_{ij}}{(p^2+\beta^2)(p'^2+\beta^2)}\,,
\end{equation}
which is a symmetric matrix in the two channels (1) $\Sigma \pi$ and
(2) $\bar{K}N$.

Define a diagonal matrix of form factors
\begin{equation}
\Pi_{ij} = \frac{1}{(p_i^2+\beta^2)}\,\delta_{ij}\,,
\end{equation}
where the momentum $p_i$ in channel $i$ is fixed in terms of the
overall c.m.\ energy $W$. For the Yamaguchi form factors of
Eq.~(\ref{Yamaguchi}), define a second diagonal matrix of dispersion
integrals:
\begin{equation}
\label{dispersion} \Delta_{ij} =
\frac{m_i}{4\pi\beta(\beta-ip_i)^2}\,\delta_{ij}\:,
\end{equation}
where $m_i$ is the reduced mass in channel $i$.

The Schr\"odinger equation can then be resolved to give the purely
$S$--wave $T$--matrix
\begin{equation}
\label{T1} T(W) =\Pi(I+A\Delta)^{-1}A\,\Pi\,.
\end{equation}

The resulting differential cross sections becomes
\begin{equation}
\left(\frac{d\sigma}{d\Omega}\right)_{\!\!j\to i} =
\frac{m_im_j}{4\pi^2}\frac{p_i}{p_j}\,\left|T_{ij}\right|^2\:.
\end{equation}

The available experimental data are fit with the input $I=0$
potentials~\cite{Gal}
\begin{equation}
A_{11} = -0.176\,\textrm{fm}^2\,,\hspace{0.5cm} A_{12} =
1.414\,\textrm{fm}^2\,,\hspace{0.5cm} A_{22} =
-1.370\,\textrm{fm}^2\,
\end{equation}
with $\beta=3.5\,$fm$^{-2}$. These values lead to a $\Lambda(1405)$
pole at $W=(1406.5-25i)$\,MeV. The relation between momenta and
overall energies was evaluated using non-relativistic kinematics,
though this might be questioned for $\pi\Sigma$.

The above formalism is suitable for the description of free coupled
$\pi\Sigma/\bar{K}N$ scattering. Suppose now that we introduce a
third channel, in this case the initial $pp$ system, that is coupled
weakly to these two. In lowest order perturbation theory, the
transition matrix element from channel--3 to the other two is then
given by
\begin{equation}
\label{T2} \mathcal{T}_{i}(W)
=\left[\Pi(I+A\Delta)^{-1}\right]_{ij}C_j\,.
\end{equation}
The $C_j$ represents a column vector of the initial preparation of
the system in the raw $\pi\Sigma/\bar{K}N$ states before the final
state interaction is introduced. In keeping with the assumption of a
short-range transition, we neglect any energy or mass dependence of
the preparation vector $C$.

The values of $|\mathcal{T}|^2$ must be multiplied by the phase
spaces for $pp\to pK^+\{pK^-\}$ and $pp\to pK^+\{\Sigma^0\pi^0\}$,
with a consistent relative normalisation. The shapes of the
distributions are determined by the (complex) ratio $C_2/C_1$. To
simplify the notation, we take $C_1=1$ and, purely for presentational
purposes, normalise each data set to the integrated measured cross
sections. We can then ask whether the relative normalisation is as
predicted.

\begin{figure}[htb]
\begin{center}
\includegraphics[clip,width=5.5cm]{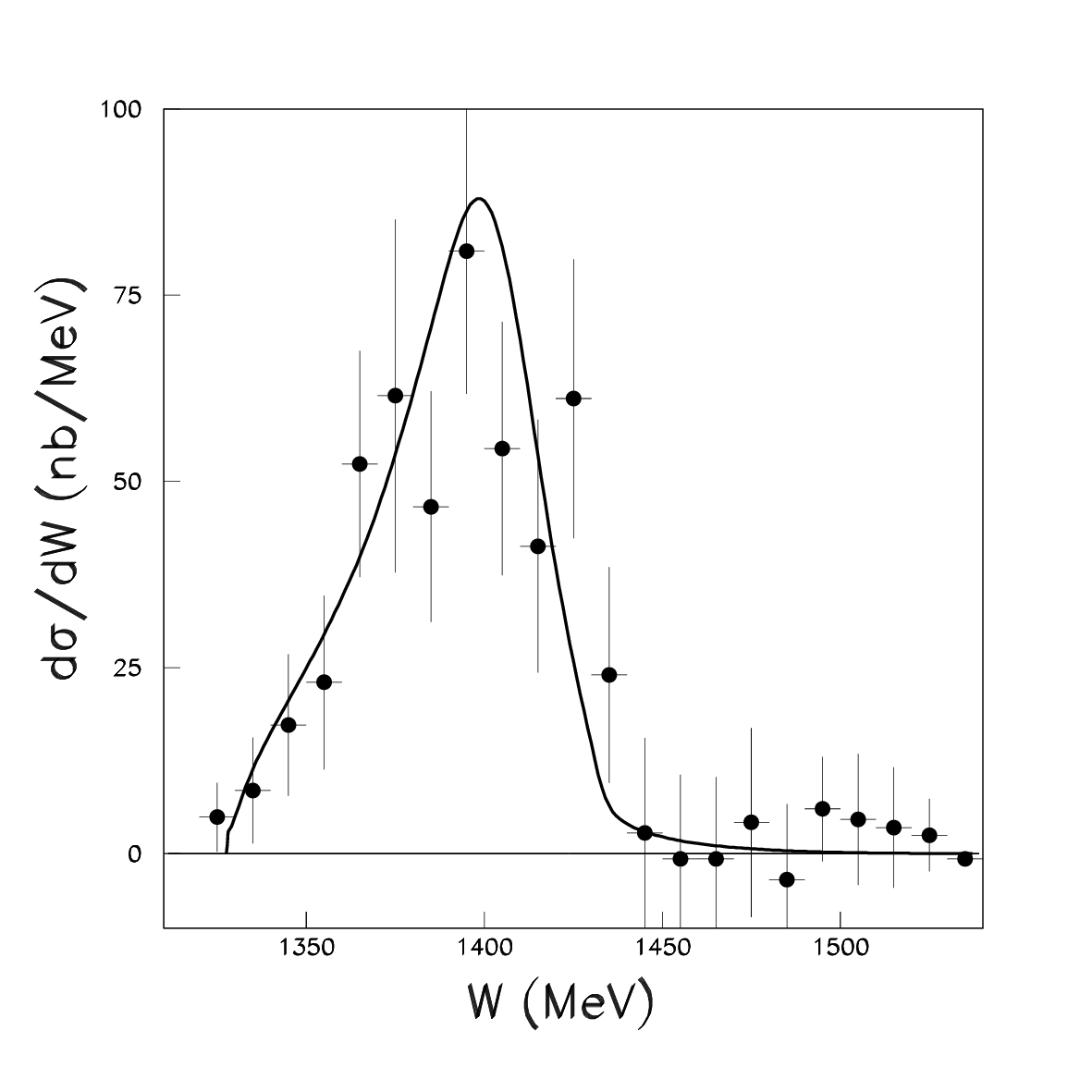}
\includegraphics[clip,width=5.5cm]{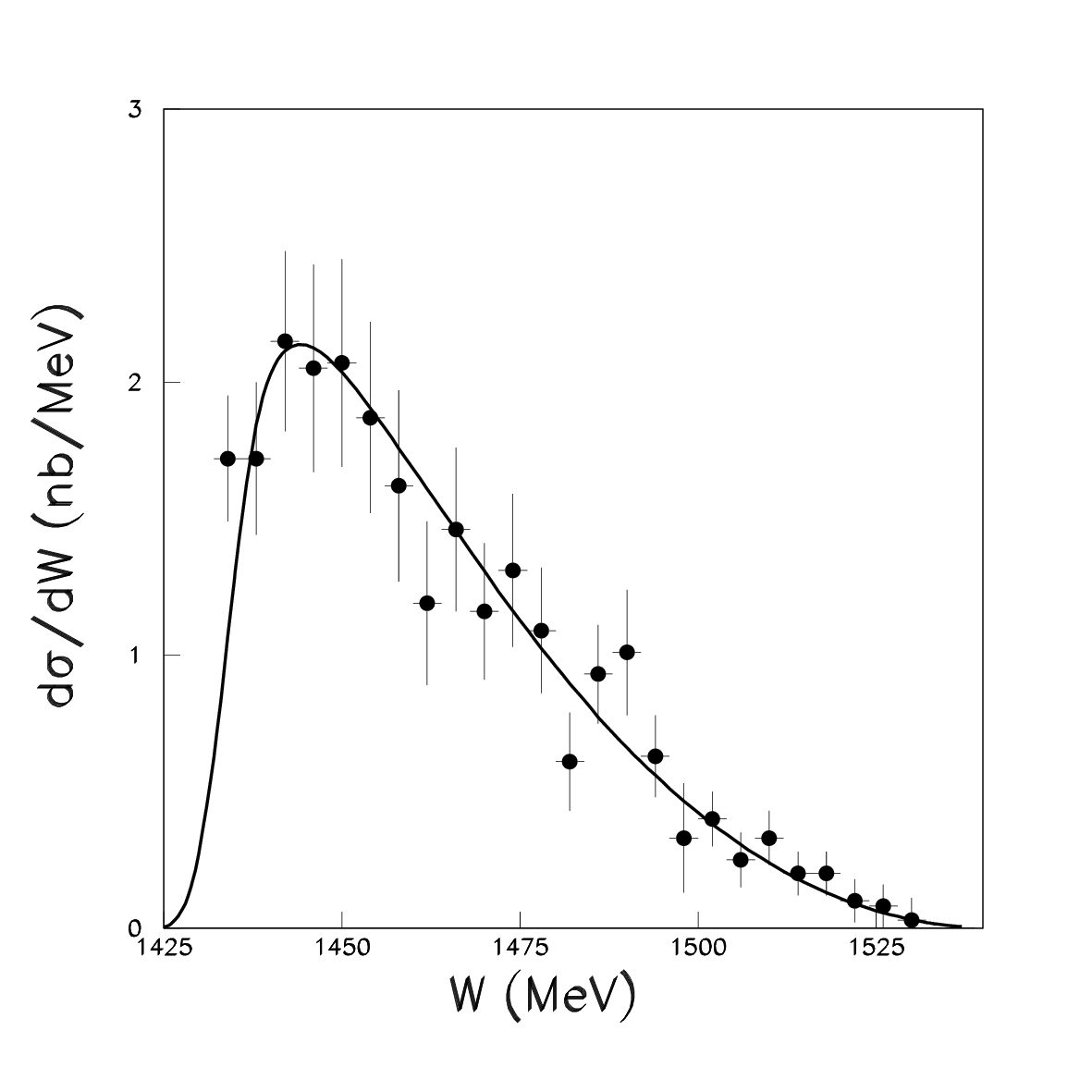}
\caption{Cross sections for (left) $pp\to
pK^+\{\Sigma^0\pi^0\}$~\cite{Iza} and (right) $pp\to
pK^+\{pK^-\}$~\cite{Maeda08} at a beam momentum of 3.65~GeV/c in 
terms of the $\Sigma^0\pi^0$ and $pK^-$ invariant masses, 
respectively. In the latter case the contribution from $\phi$ production was
excluded. Theoretical predictions in the
separable potential model were obtained with
$C=-0.7i$.\label{speculate1}}
\end{center}
\end{figure}

Figure~\ref{speculate1} is obtained if the purely imaginary value
$C_2=-0.7i$ is used. This value predicts a total cross section ratio
\begin{equation}R_{K\pi}=\sigma(pp\to pK^+\{pK^-\})/\sigma(pp\to
pK^+\{\Sigma\pi\}^0) = 9.4\times 10^{-3}\,,
\end{equation}
to be compared to the experimental value of $(22\pm 8)\times
10^{-3}$, where the contribution from $\phi$ production is not included. This
is perfectly acceptable agreement, given the model's 
simplicity. Apart from other defects, the $pp$ \emph{fsi} has been
neglected, as has any quantum mechanical interference arising from
the presence of two final protons.

There is one rather tricky point that must be mentioned. In order to
get good agreement for the shape of the $K^-p$ spectrum, Maeda
\emph{et al.}~\cite{Maeda08} needed to put in the \emph{fsi} of the
$K^-$ with \emph{both} protons. In the present approach it is assumed
that the whole $K^-p$ distribution shown does in fact come from the
$\Lambda(1405)$ channel, even though there are two protons in the
final state. This is in fact completely consistent with the
factorisation assumption that the $K^-$ can have simultaneous
\emph{fsi} with both protons.

\begin{figure}[htb]
\begin{center}
\includegraphics[clip,width=5.5cm]{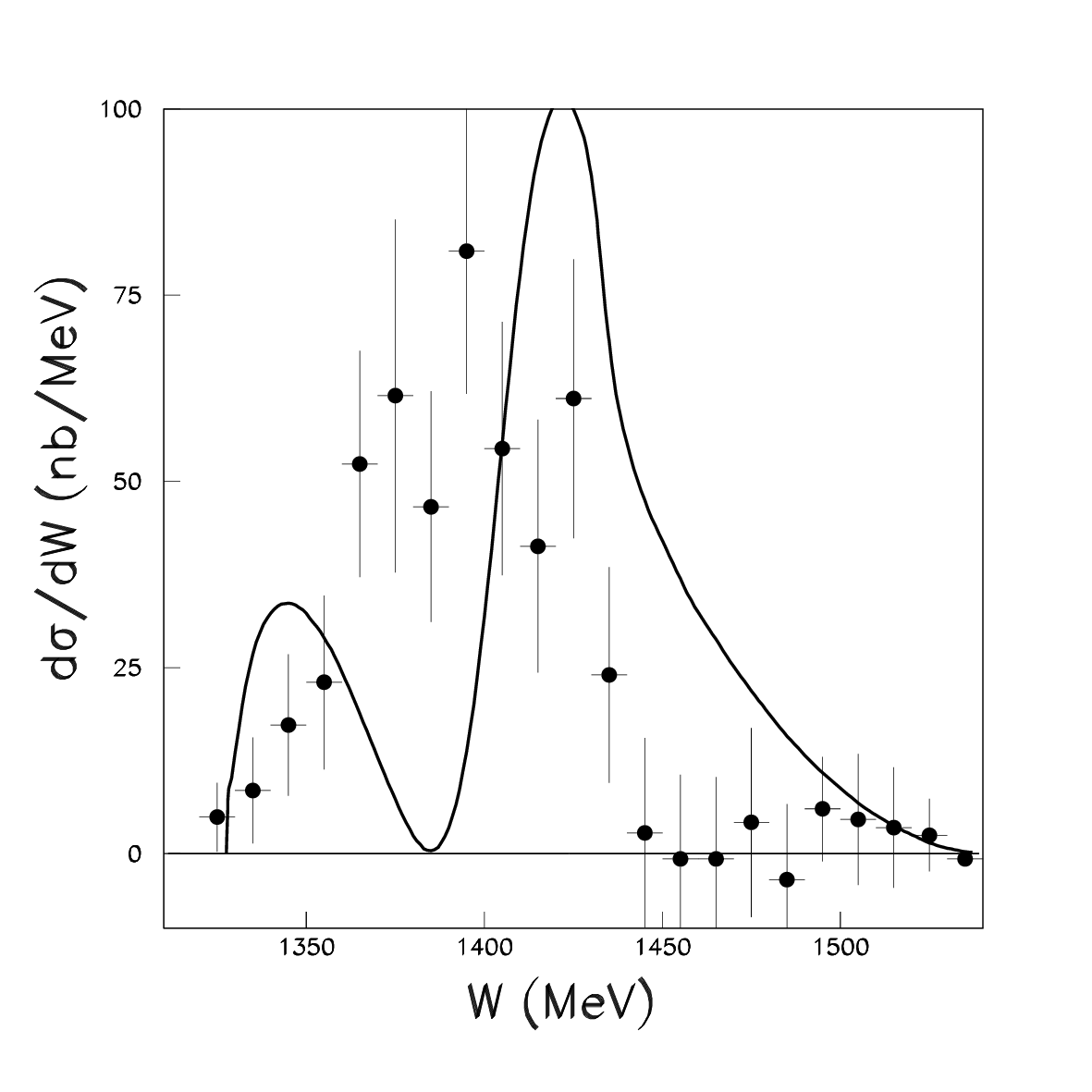}
\includegraphics[clip,width=5.5cm]{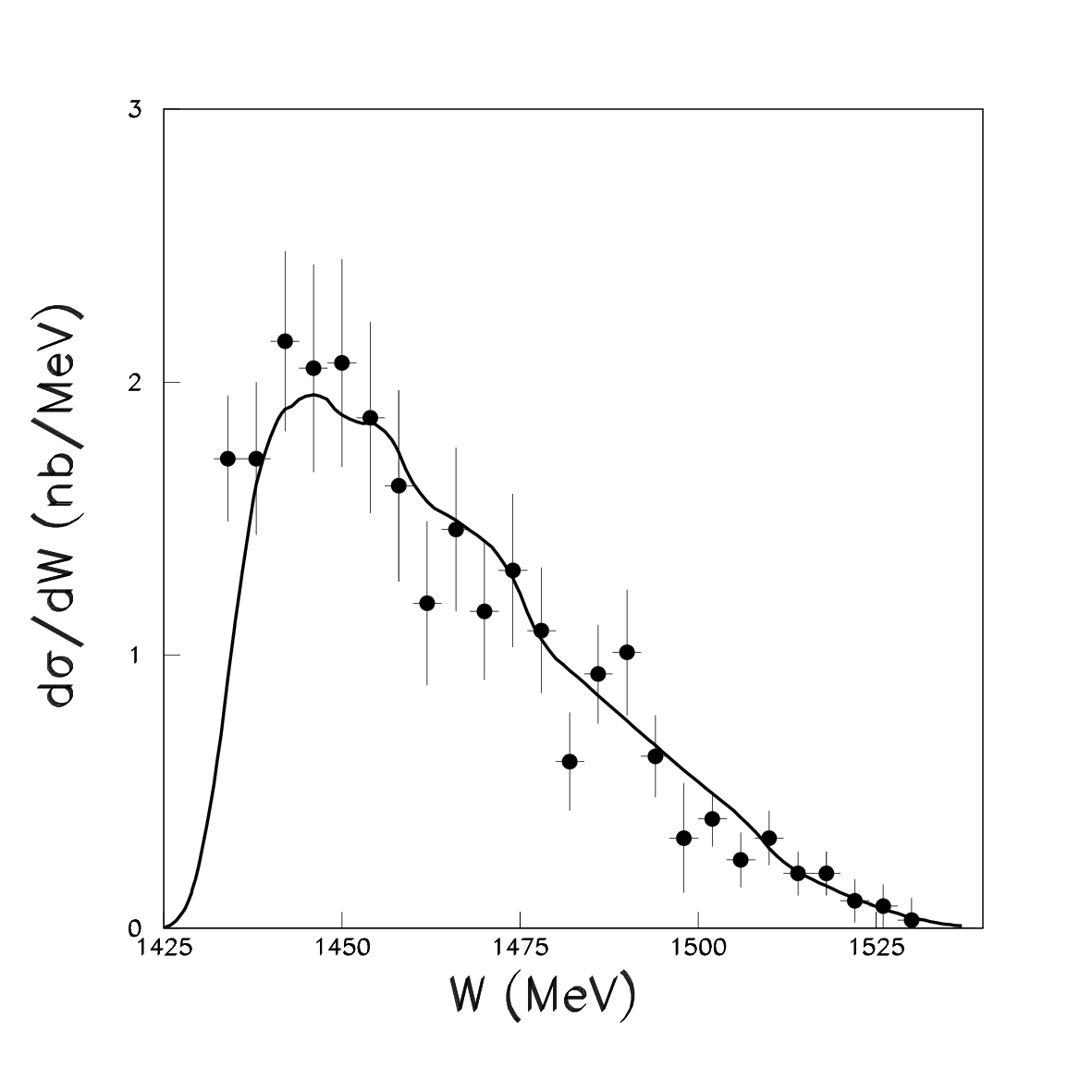}
\caption{Cross sections for (left) $pp\to
pK^+\{\Sigma^0\pi^0\}$~\cite{Iza} and (right) $pp\to
pK^+\{pK^-\}$~\cite{Maeda08}, as in Fig.~\ref{speculate1}.
Theoretical predictions were obtained with
$C=+0.5$.\label{speculate2}}
\end{center}
\end{figure}

The obvious question now is: ``How stable are the results to changes
in the value of $C_2$?''. The short answer is: ``not very!''. This is
illustrated in Fig.~\ref{speculate2}, where a purely real value
$C_2=+0.5$ is chosen. The $K^-p$ spectrum doesn't change too much
(although it looks as though my hand must have been shaking when I
drew it) but the $\Sigma\pi$ can vary enormously. For this value of
$C_2$ one can even generate a double-peaked structure. The predicted
cross section ratio of $18 \times 10^{-3}$ is close to experiment,
but that is pretty meaningless in view of the complete failure to
describe the shape.

The results presented here are still preliminary, and no attempt has
been made to include any contribution from the production of
isovector $K^-p$ pairs to the cross section. Nevertheless some
general conclusions might be drawn. The first, fairly obvious one, is
that the ratio of the $pp\to pK^+\Sigma^0\pi^0$ and $pp\to pK^+pK^-$
comes out about right in this very hand-waving approach. As a
consequence it seems likely that the same underlying reaction
mechanism drives both processes and, hence, that one should try
to estimate the two cross sections together in a realistic dynamical model.

Following from the above argument, the fact that $K\bar{K}$ scalar
resonances cannot contribute in a major way to $pp\to
pK^+\Sigma^0\pi^0$, means that they are unlikely to do so for $pp\to
pK^+pK^-$ either, though they could distort the $K^+K^-$ spectrum at
low invariant masses through a \emph{fsi}~\cite{Alexey3}.

Why is the $K^-p$ spectrum fairly stable to changes in the
preparation vector $C_i$ while the $\Sigma^0\pi^0$ distribution can
change dramatically? The origin of this probably lies in the form of
the separable potential used to describe the channel coupling.
Gal~\cite{Gal2} points out that the $\Sigma\pi$ diagonal interaction
used in Ref.~\cite{Gal} is much weaker than that of the chiral
perturbation theory approaches~\cite{Oller}. In other words, the
$\Sigma\pi$ is really being driven here more by the $K^-p$. To check
this we would really need a separable potential fitted to the chiral
perturbation amplitudes.

Finally we turn to the related question of whether the
$\Lambda(1405)$ is actually a single resonance or whether there are
two closely spaced states that might be coupled differently to
different channels. The ANKE $pp\to pK^+\Sigma^0\pi^0$ data show no
sign of any two-peak structure~\cite{Iza} but Geng and
Oset~\cite{Oset} have shown that this is not necessary or even likely 
in a two-pole scenario. It depends on the background and on how the
state is prepared. In a sense, this is also what is found here in a
much more intuitive approach. The shape of the spectra will depend
upon the preparation vector as well on as the coupling potentials. In brief,
two poles do not necessarily imply two peaks and two peaks do not necessarily
imply two poles! 

Coupled-channel effects in strange particle production at
intermediate energies seem to be a rich field for theorists to till
in the next few years provided that we are given more data,
preferably with higher statistics\,!

I should like to thank the organisers for meeting support.
Correspondence with Avraham Gal on the potential of Ref.~\cite{Gal}
proved most helpful, as did the email exchanges with Eulogio Oset and Nina
Schevchenko. I am also grateful to Alexey Dzyuba for help 
with the phase-space evaluation.

%
%

\end{document}